\documentclass[12pt,preprint,apj]{emulateapj}
\usepackage{epstopdf}
\shorttitle{ Distance to M74}
\shortauthors{Jang \& Lee}

\begin{document}

\title{
% { \bf DRAFT  \today} \\
The Tip of the Red Giant Branch Distance to the Perfect Spiral Galaxy M74 
Hosting  Three Core-Collapse Supernovae %Type II-P supernova SN 2013ej %Three Core-Collapse Supernovae %
}
 
\author{In Sung Jang and Myung Gyoon Lee }
\affil{Astronomy Program, Department of Physics and Astronomy, Seoul National University, Gwanak-gu, Seoul 151-742, Korea}
\email{isjang@astro.snu.ac.kr, mglee@astro.snu.ac.kr }

%==============================================================================================================

\begin{abstract}
M74 (NGC 628) is a famous face-on spiral galaxy, hosting three core-collapse supernovae (SNe):SN Ic 2002ap, SN II-P 2003gd, and SN II-P 2013ej. However its distance is not well known.
We present  distance estimation for this galaxy based on the Tip of the Red Giant Branch (TRGB) method.  
We obtain photometry of the resolved stars in the arm-free region of M74 from $F555W$ and $F814W$ images in the Hubble Space Telescope archive.
The color-magnitude diagram of the resolved stars  
shows a dominant red giant branch (RGB) as well as blue main sequence stars, red helium burning stars, and asymptotic giant branch stars.
The $I$-band luminosity function of the RGB stars shows the TRGB to be at 
$I_{TRGB}= 26.13\pm0.03$ mag, and $T_{RGB}=25.97\pm0.03$.
From this we derive the distance modulus to M74 to be
 $30.04 \pm 0.04 {\rm (random)} \pm 0.12 {\rm (systematic)}$ 
 (corresponding to a linear distance,   $10.19\pm 0.14 \pm0.56$ Mpc).
 With this distance estimate, we calibrate the standardized candle method for SNe II-P.
From the absolute magnitudes of SN 2003gd, we derive a value of the Hubble constant, $H_0 = 72\pm 6$ (random)$\pm 7$ (systematic) km s$^{-1}$ Mpc$^{-1}$. It is similar to recent estimates based on the luminosity calibration of Type Ia supernovae.
\end{abstract}


\section{Introduction}

M74 (NGC 628) is often called the perfect spiral galaxy, playing as a prototype of SA(c).  
It hosts three core-collapse supernovae (SNe): SN Ic 2002ap, SN II-P 2003gd, and SN II-P 2013ej \citep{maz02,maz07,hen05,kim13,val14}.   
Progenitors of core-collapse SNe are one of the critical issues in the study of evolution of massive stars. Progenitors of Type II-P SNe are considered to be massive red supergiant stars with 8 -- 16 $M_\odot$ \citep{sma09,fra14}. SN II-P 2003gd in M74 has been used as one of the gold set to estimate the mass and luminosity of the progenitor of the  type II-P SNe \citep{sma09}. \\
Another Type II-P SN, 2013ej, was recently discovered in M74. It was found when it was younger than one day after explosion so that it is an important probe to understand the early evolution of Type II-P SNe \citep{kim13}.
On the other hand, Type II-P SNe have been used as powerful extragalactic distance indicators \citep{ham02,oli10}. Although Type II-P SNe are not as bright as Type Ia SNe and they
are not as precise as Type Ia SNe, they are still useful as a distance indicator for distant galaxies \citep{ham02,hen05,oli10,bos14}.
However, the current calibration of type II-P SNe are based on only a few nearby galaxies having a large uncertainty \citep{ham04,oli10}.  

M74 is one of the nearest galaxies that host Type II-P SNe and is the only nearby galaxy hosting two known Type II-P SNe, so it is an excellent galaxy for studying the properties of Type II-P SNe and to improve the calibration of Type II-P SNe as a distance indicator.
However, the distance to M74 is still not well known, with previous estimates ranging from 7 Mpc to 10 Mpc \citep{sha96, hen05, her08,oli10,gus14}.
In this study, we present distance estimation of this galaxy based on the Tip of the Red Giant Branch (TRGB) method, known to be a precise standard candle \citep{lee93,fre10}.  

\section{Data and Data Reduction}

%%%%%%%%%%%%%%%%%%%%%%%%%%%%%%%%%%%%%%%%%%%
%% TABLES
%%%%%%%%%%%%%%%%%%%%%%%%%%%%%%%%%
\begin{deluxetable*}{lcccccc}
\tabletypesize{\footnotesize}
%\tabletypesize{\scriptsize}
%\tabletypesize{\tiny}
\setlength{\tabcolsep}{0.05in}
%\rotate
\tablecaption{A Summary of $HST$/ACS Observations  of M74 in the Archive}
\tablewidth{0pt}

\tablehead{ \colhead{Field} & \colhead{R.A.(2000)} &  \colhead{Dec(2000)} & \colhead{Prop ID.} & \colhead{Filter} & \colhead{Exposure} %& \colhead{File name}
}
\startdata
F1 	& $01^h36^m52.^s06$	& $15\arcdeg45\arcmin 50\farcs8$	& 10402 & F814W	& 360 s	\\%& j96r22rcq$\_$flc.fits	 \\
	&  				& 				&		& F814W	& 360 s	\\%& j96r22req$\_$flc.fits	 \\
\hline
F2 	& $01^h36^m51.^s10$	& $15\arcdeg45\arcmin47\farcs0$	& 9796  & F555W	& 530 s	\\%& j8ol08edq$\_$flc.fits	\\
	&  				& 				&		& F555W	& 530 s	\\%& j8ol08efq$\_$flc.fits	 \\
	&  				& 				&		& F555W	& 530 s	\\%& j8ol08eiq$\_$flc.fits	 \\
	&  				& 				&		& F555W	& 530 s	\\%& j8ol08emq$\_$flc.fits	 \\
\hline
F3 	& $01^h36^m47.^s48$	& $15\arcdeg45\arcmin28\farcs3$	& 9796  & F555W	& 550 s	\\%& j8ol04f1q$\_$flc.fits \\
	&  				& 				&		& F555W	& 550 s	\\%& j8ol04flq$\_$flc.fits	 \\
	&  				& 				&		& F555W	& 550 s	\\%& j8ol04g6q$\_$flc.fits	 \\
	&  				& 				&		& F555W	& 550 s	\\%& j8ol04gsq$\_$flc.fits	 \\
	\hline
F3 	& $01^h36^m47.^s48$	& $15\arcdeg45\arcmin28\farcs3$	& 9796& F814W	& 390 s	\\%& j8ol04f0q$\_$flc.fits	 \\
	&  				& 				&		& F814W	& 390 s	\\%& j8ol04fjq$\_$flc.fits 	 \\
	&  				& 				&		& F814W	& 390 s	\\%& j8ol04g4q$\_$flc.fits	 \\
	&  				& 				&		& F814W	& 390 s	\\%& j8ol04gqq$\_$flc.fits	 \\
\hline
\multicolumn{4}{l}{Integrated exposure times for the overlapped regions} & F555W & 4320 s \\%& final.555.drc.fits\\
\multicolumn{4}{l}{ } & F814W & 2280 s \\%& final.814.drc.fits\\
\hline
\enddata
\end{deluxetable*}

\begin{figure}
\centering
\includegraphics[scale=0.85]{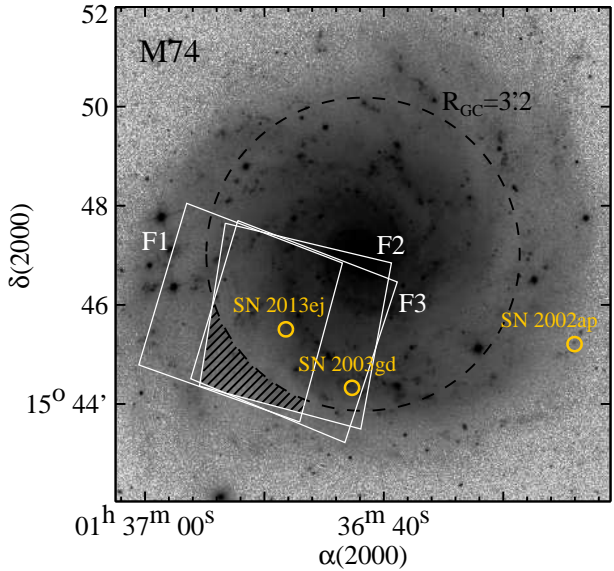} %finding.eps} %white.eps}
\caption{Finding chart for $HST$/ACS fields overlaid on the grayscale map of a $10\arcmin \times 10\arcmin$ SDSS $r$-band image of M74. 
%The hatched region with denotes the region we used for our TRGB distance estimation. 
For the TRGB analysis we only used resolved stars located in the overlapping regions at $R_{GC} \ge 3\farcm2$ indicated by the hatched lines. %were used in  
Positions of three known SNe are marked by open circles.
}
\label{finding}
\end{figure}

We used Hubble Space Telescope ($HST$)/Advanced Camera Survey (ACS) images of M74  in the archive as listed in Table 1. 
There are three overlapping $HST$ fields, marked in Figure \ref{finding}, covering the southeast region of M74, which displays a grayscale map of Sloan Digital Sky Survey (SDSS) $r$-band image of M74. 
We also marked the positions of the three known SNe in M74 in this figure.

\begin{figure*}
\centering
\includegraphics[scale=0.75]{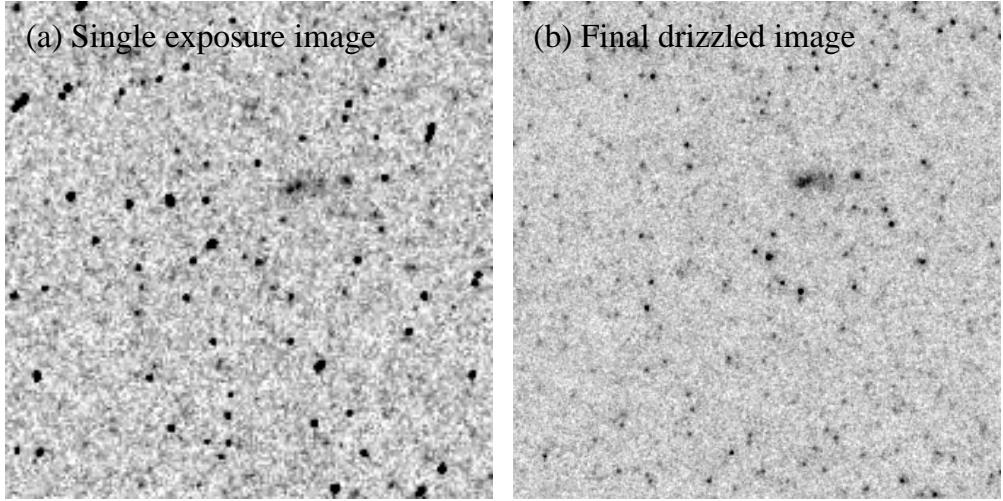} %{resolve.eps} %white.eps}
\caption{(a) Grayscale map of the 
distortion-corrected single $F814W$ image with a pixel scale of $0\farcs05$ and exposure time of 390 s for a $10\arcsec \times 10\arcsec$ section in the $HST$ field.
A significant fraction of the bright  sources are cosmic rays.
(b) Grayscale map of the reconstructed $F814W$ image for the same region, with a resampling of $0\farcs03$ pixel scale and total exposure time of 2280 s. % from the six separate images using AstroDrizzle.
Cosmic rays were successfully removed and faint stellar objects are better seen.
}
\label{resolve}
\end{figure*}

Single exposure images were obtained with short exposure times ranging from 360 s to 550 s so that it is difficult to derive photometry of resolved red giant stars from them. 
We constructed deeper drizzled images for each filter combining the flat field-calibrated single images corrected for  charge transfer efficiency (indicated by $\_$flc.fits images)
%flat fielded and charge transfer efficiency corrected single images (indicated by $\_$flc.fits images) 
using Tweakreg and AstroDrizzle task in DrizzlePac provided by the Space Telescope Science Institute (STScI) 
  (http://www.stsci.edu/hst/HST$\_$overview/drizzlepac/).
We adopted a PIXFRAC value of 0.7 and created final images with
a pixel scale, $0\farcs03$ per pixel.
Total exposure times are 4320 s for $F555W$ image and 2280 s for $F814W$ image, respectively. 
Both images show full width at half-maximum (FWHM) of 2.89 pixels, corresponding to $0\farcs$087.
Figure \ref{resolve} displays a $10\arcsec \times 10\arcsec$ section of a single $F814W$ image and that of the drizzled image reconstructed from six single images. % are shown in . 
It is seen that the reconstructed image is free from cosmic rays and superior to the single image in showing faint objects.

Instrumental magnitudes of the point sources were derived using the standard routine, IRAF/DAOPHOT \citep{ste94}. 
Point spread function (PSF) images were constructed based on $\sim 30$ bright and isolated point sources located in a relatively less crowded region at the galactocentric distance $R_{GC} \ge 3\farcm2$.
We used $2.5\sigma$ as a detection threshold, and ran a sequence of DAOFIND/PHOT/ALLSTAR tasks.
Aperture correction values  were calculated by checking the curve of growth of the point sources to the aperture radius of $0\farcs5$ (16.67 pixels). Corresponding values are $-0.14\pm0.03$ for $F555W$ and $-0.19\pm0.02$ for $F814W$, respectively. For additional correction values to infinity, we used revised values of --0.099 for $F555W$ and --0.098 for $F814W$ as recommended by STScI.
Instrumental magnitudes were converted to Johnson-Cousins  $VI$ magnitudes using the equations described in \citet{sir05}.
The average errors for this transformation are 0.02 mag.

\section{Results}

\subsection{Color-Magnitude Diagrams} 

\begin{figure*}
\centering
\includegraphics[scale=0.8]{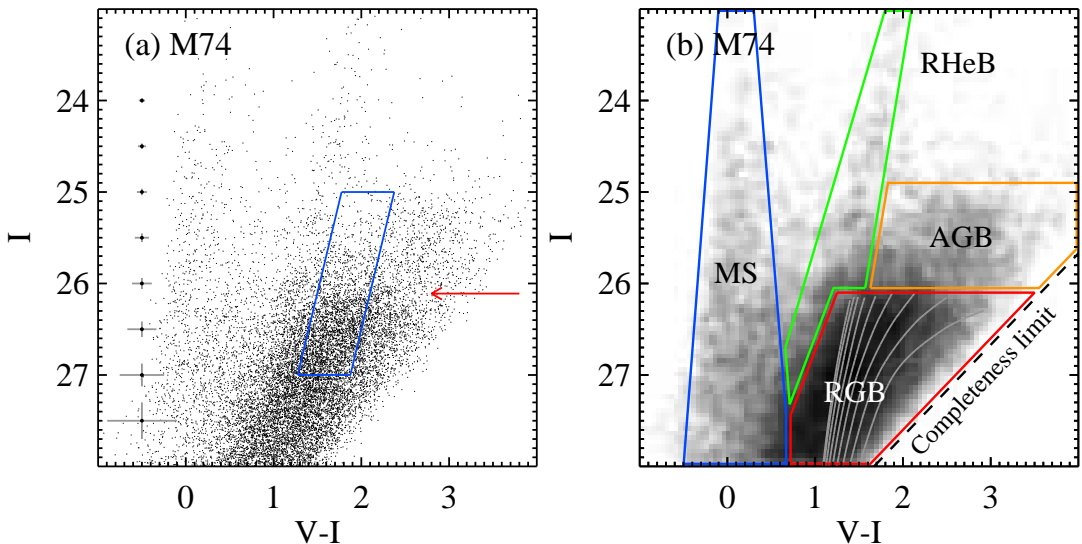} %{cmd.eps} %white.eps}
\caption{(a)  $I-(V-I)$ color-magnitude diagram of the  resolved stars in the selected region of M74. Only the stars inside the slanted box were used for the TRGB analysis.
(b) A Hess diagram for resolved stars in the same region as that of panel (a).
Major populations of stars are indicated:
MS for massive main sequence stars, RHeB for red helium burning stars, 
AGB for asymptotic giant branch stars, and  RGB for red giant branch stars.
Curved lines in the RGB domain represent the 12 Gyr stellar isochrones 
for [Fe/H] = $-0.6 \sim -2.2$ in steps of 0.2 provided by the Dartmouth group \citep{dot08}.
%(b) The luminosity function of the selected RGB stars (solid line) and the corresponding edge detection response (dashed line). %based on the method described in \citet{sak96} (dashed line). 
%Note the peak of the edge detection response show the TRGB to be at $I = 26.06 \pm 0.04$.
}
\label{cmd}
\end{figure*}

M74 is an almost face-on spiral galaxy and the $HST$/ACS fields covered disk regions with spiral arms. 
For the TRGB analysis we chose a $R_{GC} \geq 3\farcm2$ region, marked by the hatched region in Figure \ref{finding}, with the lowest sky background level, avoiding arms and star-forming regions.
The foreground reddening value for M74 
is known to be small, $E(B-V)=0.062$  \citep[NASA/IPAC Extragalactic Database]{sch11}. 
This corresponds to $A_I=0.106$ and $E(V-I)=0.086$.  

Figures \ref{cmd} (a) and (b) display color-magnitude diagrams (CMDs) of the resolved stars in the selected region.
It is expected that the contamination due to galactic foreground stars is negligible in this figure, because the galactic coordinates of M74 are $l=138.6$ deg and $b=-45.7$ deg and the field of view of the selected region is small. 
A few features are noted in this figure as follows. First, there is a vertical feature at $(V-I)\sim 0.0$. This is a blue plume consisting of massive main-sequence stars and blue supergiants in the disk of M74.
Second, a slanted feature reaching to $I\sim 23.0$ at $(V-I)\sim2.0$ is seen there. This corresponds to red supergiants and core-He-burning massive stars, which are also located in the disk
of M74.
Third, the most distinguishable feature is a broad slanted feature extending to $I\sim 26.1$ mag at $(V-I)\sim2.0$, including the largest amount of stars
in the figure. It is a red giant branch (RGB). The stars in this RGB are probably old stars in M74. 
This RGB is approximately matched by 12 Gyr stellar isochrones 
for [Fe/H] = $-0.6 \sim -2.2$ provided by the Dartmouth group \citep{dot08},
as shown in Figure \ref{cmd}(b).
The stars located above the RGB are mostly AGB stars.

\subsection{Distance Estimation}

\begin{figure*}
\centering
\includegraphics[scale=0.8]{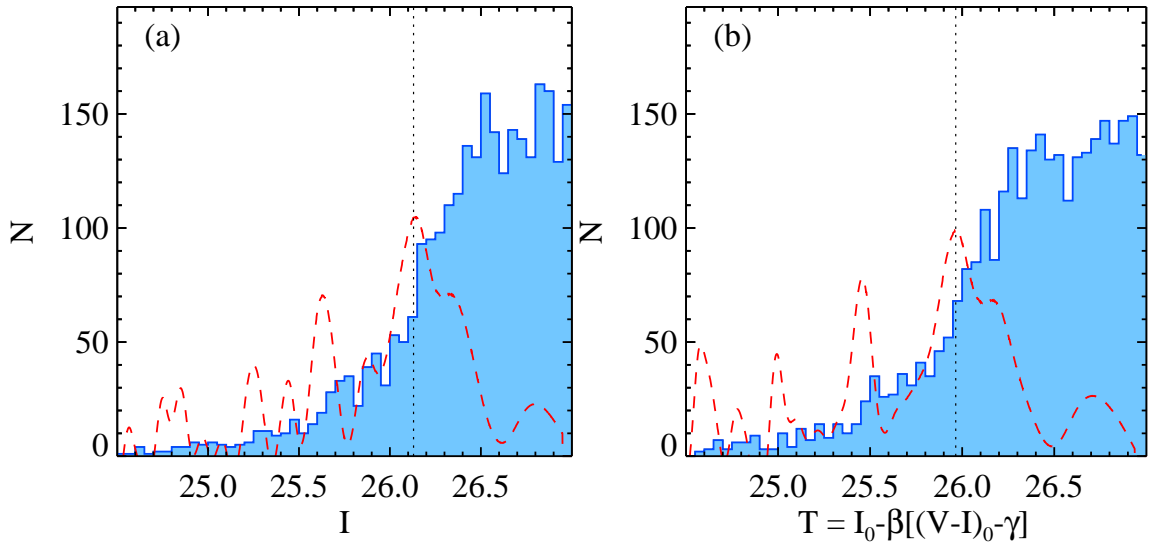} %{edres.eps} %white.eps}
\caption{$I$-band luminosity function (a) and T magnitude luminosity function (b) of the resolved stars at $R_{GC} \ge 3\farcm2$ (filled histogram) and corresponding edge-detection response (dashed line). Prominent peaks corresponding to the TRGB are measured to be at $I_{TRGB} = 26.13\pm0.03$ and $T_{RGB} = 25.97\pm0.03$. 
%(b) A distribution of edge-detection responses with respect to the galactocentric radii. 
%White is strong, black is weak response.
%The strongest peak is seen at $26.05 \sim I \sim 26.10$, 
%regardless of galactocentric radii.
}
\label{edres}
\end{figure*}

We  estimated the distance to M74 from the photometry of the resolved stars using the TRGB method, as described in \citet{lee93}, \citet{lee12}. 
First, we derived the $I$-band luminosity function of the red giants, counting the stars inside the box marked in Figure \ref{cmd} (a).
We selected this box region to derive the luminosity function of relatively metal-poor red giants because of three reasons. 
First, the bluer (metal poorer) RGB is less curved than the redder (metal richer) RGB in the CMD. 
Second, the bluer RGB is less affected by the photometric incompleteness than the redder RGB. 
Third, the bluer RGB is expected to be less affected by the internal reddening than the redder RGB.
 
The field used in this study is fairly close to the disk of M74, so there is a possibility that internal extinction may affect our distance estimation.
Since the field being investigated is squarely in the disk of M74, stars must be seen on both sides of the disk so at the very least half of the TRGB population must be affected by disk reddening.
Therefore the observed RGB is expected to include stars with and without reddening.
The bluest part of the observed RGB represents the unreddened stars, while the red part of the RGB contains both unreddened red stars and reddened blue stars.
Therefore we used the blue part of the RGB, as shown by the slanted box in Figure \ref{cmd} (a), to avoid the internal reddening effect as much as possible.

Figure \ref{edres} (a) displays the derived $I$-band luminosity function (filled histogram) and edge-detection response (dashed line) of the stars located inside the slanted box in Figure \ref{cmd} (a).
The luminosity function of the RGB stars shows a clear jump at $I \sim 26.1$, corresponding to the TRGB. 
We derive a value of $I_{TRGB}= 26.13\pm0.03$, from a quantitative analysis of $\sim 3100$ resolved stars  using the edge-detecting algorithm \citep{sak96}. %,men02,mou10}.
The edge-detection response function for given luminosity function $N(I)$ is described by 
$E(I)$ ($= N (I+ \sigma_I ) -   N (I - \sigma_I )  $) where $\sigma_I$ is the mean photometric error. 
The value of the TRGB magnitude was determined from the peak value of the edge-detection response function. %, as plotted in Figure \ref{edres} (a).
\citet{mad09} noted that at least 400--500 stars in the one-magnitude interval below the TRGB level are required to detect the TRGB within $\pm0.1$ mag statistical uncertainty.
The number of stars in that magnitude range used in this study is $\sim2600$, which is large enough for precise determination of the TRGB magnitude.
% Thus our measurement is statistically stable.
We obtained the median color value of the TRGB  from the color of the selected brightest RGB stars: 
 $(V-I)_{\rm TRGB} = 1.82\pm0.02$.
Errors of $I_{TRGB}$ and $(V-I)_{\rm TRGB}$ values were estimated from the bootstrap resampling method with ten 10,000 simulations.

By applying the TRGB calibration described in \citet{riz07}
($M_{{\rm I,TRGB}} = -4.05 + 0.217[ (V-I)_{0, TRGB} -1.6]$),
we obtained a value for the distance modulus of
$30.04 \pm 0.03 ({\rm random}) \pm 0.12 ({\rm systematic})$.
%(corresponding to a linear distance,   $10.2\pm 0.1 \pm0.5$ Mpc).
We derived a value of the systematic error  to be 0.12, from the combination of
the TRGB calibration error (0.12), aperture correction error (0.02), standard transformation error (0.02), and galactic extinction error (0.01, which is 10\% of galactic extinction).\\

We also applied a modified TRGB detection method introduced by \citet{mad09}.
  This method introduces the $T$ magnitude ($T = I_0 - \beta[(V-I)_0 - \gamma]$, where $\beta=0.20\pm0.05$, $\gamma=1.5$)
 to make the discontinuity at the TRGB sharper. 
 The absolute $T$ magnitude of the TRGB is --4.05.
We applied this method to the same RGB star sample, 
deriving $T_{RGB}=25.97\pm0.03$, as shown in Figure \ref{edres} (b). 
From this we obtained a distance modulus, 
$(m-M)_0 = T_{RGB} + 4.05 = 30.02\pm0.03\pm0.12$. 
This value is consistent with the distance modulus 
derived from the $I$-band magnitudes, $30.04\pm0.03$.
We adopt $(m-M)_0 = 30.04\pm0.03\pm0.12$ (corresponding to a linear distance,   $10.19\pm 0.14 \pm0.56$ Mpc) as the distance modulus to M74 to be consistent with our results for other galaxies.
Our distance estimates based on these two methods  are summarized in Table \ref{dist_summary}.

\begin{deluxetable*}{lcc}
\tabletypesize{\footnotesize} % \footnotesize \scriptsize
%\tabletypesize{\tiny}
\setlength{\tabcolsep}{0.05in}
%\rotate
\tablecaption{A Summary of TRGB Distance Measurements for M74}
\tablewidth{0pt}
%\tablehead{ \multirow{2}{*}{Parameter} & \multicolumn{2}{c}{TRGB calibration} \\
%\hline 
% & \colhead{\citet{riz07}} & \colhead{\citet{mad09}}   } 
\tablehead{ \colhead{Parameter} &  \colhead{$I$~ magnitude} & \colhead{$T$~ magnitude} }
\startdata
%TRGB magnitude, $F814W_{TRGB}$				& $0\pm0.0$  \\
%Median TRGB color, $(F555W-F814W)_{TRGB}$		& $0\pm0.0$  \\
TRGB magnitude					& $I_{TRGB} = 26.13\pm0.03$ & $T_{RGB} =  25.97\pm0.03$\\
%T magnitude, $T$					& $25.97\pm0.03$  \\
Median TRGB color, $(V-I)_{TRGB}$				& $1.82\pm0.02$  & ...\\
%Foreground extinction at $V$, $A_{V,Gal}$	& 0.192 \\ % \pm 0.019
%Foreground extinction at $I$, $A_{I,Gal}$	& 0.106 \\ % \pm 0.011
%Foreground reddening, $E(V-I)_{Gal}$		& 0.086 \\ % \pm 0.022
Intrinsic TRGB magnitude, $I_{0,TRGB}$ 		& $26.02\pm0.03$ & ...\\
Intrinsic mean TRGB color, $(V-I)_{0,TRGB}$	& $1.73\pm0.02$ & ...\\
Absolute TRGB magnitude, $M_{I,TRGB}$		& $-4.02$ & $-4.05$\\
Distance modulus, $(m-M)_0$					& $30.04\pm0.03\pm0.12$ & $30.02\pm0.03\pm0.12$\\
Distance					& $10.19\pm0.14 \pm0.56$ Mpc & $10.09\pm0.14 \pm0.56$ Mpc \\
\hline
%\tablenotetext{a}{$(m-M)_{0,M66}=30.12\pm0.03$ (Lee \& Jang 2013)}
\enddata
\label{dist_summary}
\end{deluxetable*}

\section{Discussion}

\subsection{Comparison with Previous Distance Estimates}

\begin{figure}
\centering
\includegraphics[scale=0.8]{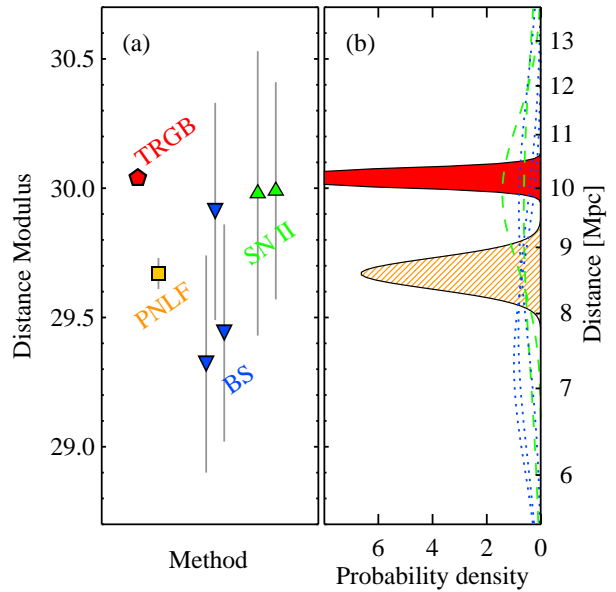} %{distance.eps} %white.eps}
\caption{(a) Comparison of distance estimates to M74 based on the TRGB (pentagon) in this study, 
the PNLF (square), %Planetary Nebula Luminosity Function (square), 
the brightest stars (downward triangle), 
and the Type II-P SNe (upward triangle) in the previous studies. 
(b) Corresponding probability density curves for each measurement with a normalized Gaussian function centered at the distance modulus value with a width equal to the measurement error.
Gaussians with filled area, hatched area, dotted lines, and dashed lines represent the distance estimates based on the TRGB, the PNLF, the brightest stars, and the Type II-P SNe, respectively.
%\caption{(a) A comparison of distance estimates to M74 based on TRGB (pentagon; This study), Planetary Nebula Luminosity Function (square; \citet{her08}), Brightest Star (downward triangle; \citet{sha96, hen05}), and Type II-P Supernova (upward triangle; \citet{hen05,oli10}). (b) Corresponding probability density curves
%A probability density curve for each measurement was derived from a Gaussian function centered at the distance modulus value with a width equal to the measurement error
}
\label{distance}
\end{figure}

We compare our results with those in previous studies.
Figure \ref{distance} shows a comparison of distance measurements for M74 in this study and previous studies based on the brightest stars, planetary nebula luminosity functions (PNLFs), and SNe II-P standard candle method
\citep{sha96,her08, hen05}.
We derived a probability density curve for each measurement with a normalized Gaussian function centered at the distance modulus value with a width equal to the measurement error.
 
Our distance estimate from the TRGB method, 
$30.04 \pm 0.03 {\rm(random)} \pm 0.12 {\rm(systematic)}$,
%(corresponding to a linear distance,   $9.8\pm 0.2 \pm0.5$ Mpc)
is compared with previous estimates:
$29.32 \pm 0.40$ \citep{sha96},
$29.44 \pm 0.48$ to $29.91 \pm 0.49$ \citep{hen05} from the brightest supergiants (depending on the calibrations of the brightest supergiants), 
$29.67^{+0.06}_{-0.07}$ from the PNLFs \citep{her08}, and
$29.91 \pm 0.63$ \citep{hen05} and $29.98 \pm 0.28 $ %with a systematic error of 0.4 to 0.5 
\citep{oli10} from the SN II-P standardized candle method  applied to SN 2003gd.
Thus our estimate is larger than those based
on the brightest stars and the PNLF.
Our value happens to be similar to those derived from the SN II-P standardized candle method. 
However, the error of our estimate is much smaller than the latter. 

\subsection{Progenitors of SN2003gd and SN2013ej}

It is generally considered that the progenitors of core-collapse SNe have a minimum mass limit of 8 M$_\odot$ and
a maximum mass limit of 16 M$_\odot$ \citep{sma09,fra14}.
Two II-P SNe in M74, SN 2003gd and SN 2013ej are very useful for the study
of the progenitors of core-collapse SNe.
\citet{hen05} suggested that the progenitor of SN 2003gd is $8_{-2}^{+4}$
M$_\odot$, adopting a distance of $9.3\pm1.8$ Mpc. Thus
the progenitor mass of SN 2003gd is close to the minimum limit.
Recently \citet{fra14} estimated that the mass of the progenitor of SN 2013ej is between 8 and 16 M$_\odot$, assuming a distance of  $9.1\pm1.0$ Mpc.
The distance estimate derived in this study is 10 -- 12\% larger than those
adopted in \citet{hen05} and \citet{fra14}, slightly increasing the progenitor mass values of these SNe. 

\subsection{Calibration of Type II-P Standardized Candle Method}

\begin{deluxetable*}{lccc}
\tabletypesize{\footnotesize} % \footnotesize \scriptsize
%\tabletypesize{\tiny}
\setlength{\tabcolsep}{0.05in}
%\rotate
\tablecaption{A Summary of Absolute Calibration of SNe II-P and the Hubble Constant}
\tablewidth{0pt}
\tablehead{ \colhead{Parameter} & \colhead{SN 2003gd in M74} & \colhead{SN 2004dj in NGC 2403} & \colhead{SN 1999em in NGC 1637}}
\startdata

$V ^a$   	& $13.965\pm0.036$	& $12.046\pm0.035$ & $13.998\pm0.023$\\
$V-I^a$     & $0.786\pm0.019$	& $0.621\pm0.043$ & $0.752\pm0.019$\\
$v_{FeII}$ $^{a}$ [km/s]	& $2976\pm230$ 	& $2725\pm202$ & $2727\pm56$\\
$(m-M)_0$	& $30.04\pm0.03$ 	& $27.52\pm0.02^b$ & $30.34\pm0.19^c$\\
$M_{B}$		& $-17.72\pm0.19$	& $-17.13\pm0.21$ & $-17.94\pm0.20$\\
$M_{V}$		& $-18.08\pm0.15$	& $-17.32\pm0.16$ & $-18.41\pm0.20$\\
$M_{I}$		& $-17.94\pm0.13$	& $-17.17\pm0.12$ & $-18.24\pm0.19$\\
$H_0 (B)$ [km/s/Mpc]		& $71.5\pm7.1$		& $ 93.9\pm10.3$ & $ 64.7\pm6.8$\\
$H_0 (V)$ [km/s/Mpc]		& $72.4\pm5.9$		& $102.7\pm8.6$  & $ 62.2\pm6.1$\\
$H_0 (I)$ [km/s/Mpc]		& $72.3\pm4.9$		& $102.9\pm6.9$  & $ 62.9\pm6.0$\\
\hline
\tablenotetext{a}{The listed data are photometric and physical properties of SNe at the time of day --30 based on the transition phase, given by \citet{oli10}. }
\tablenotetext{b}{We adopted the TRGB distance to NGC 2403 of $27.52\pm0.02$ from \citet{dal09}.}
\tablenotetext{c}{We adopted the Cepheid distance to NGC 1637 of $30.34\pm0.19$ from \citet{leo02}.}
\enddata
\label{luminosity}
\end{deluxetable*}

\citet{ham02} found
a strong correlation between the expansion velocity  of SNe II-P and their bolometric luminosity at the plateau phase, showing that SNe II-P can be used as a standardized candle for distance estimation to levels of 0.4 and 0.3 mag in the $V$
and $I$-bands, respectively.
Later \citet{ham04} provided a calibration of the SNe II-P method using the four SN II-P-hosting
galaxies with Cepheid distances:
SN 1968L, SN 1970G, SN 1973R and SN 1999em. However, only SN 1999em among these has modern CCD photometry, while the others only have old photographic photometry.
\citet{oli10} presented an extensive study of the standardized candle method applied to modern data of 37 nearby SNe II-P at $z<0.06$.
Unfortunately, their calibration had to rely on only two galaxies with Cepheid distances,
SN 1999em in NGC 1637 and SN 2004dj in NGC 2403. However, the corrected absolute magnitudes of these two SNe (in their Table 7) showed large differences, 0.86, 1.13, and 1.11 mag
for the $B$, $V$, and $I$ bands, respectively.

 Here we present the first calibration of the SN II-P standard candle method based on the TRGB distances to SNe II-P host galaxies. Among the 37 nearby SNe II-P listed in \citet{oli10}, two SNe -- SN 2003gd in M74 and SN 2004dj in NGC 2403 --
%- M74 hosting SN 2003gd and SN 2013ej, and NGC 2403 hosting SN 2004dj - 
are  available for the calibration. 
Table \ref{luminosity} summarizes the results of our analysis.
We used $V$-band magnitudes, $V-I$ colors, and expansion velocities measured from the Fe II $\lambda$5169 line for each SN provided by \citet{oli10}. 
These values were corrected for Galactic extinction and interpolated at the time of day --30 based on the transition phase. 
We adopted 
%the TRGB distance to M74 from this study, and 
the TRGB distance to NGC 2403 from the estimate in \citet{dal09},
$(m-M)_0=27.52\pm0.02$, which is in good agreement
with the Cepheid distance, $(m-M)_0=27.48\pm0.24$ \citep{fre01}.
We calculated absolute magnitudes for each band and corresponding Hubble constants using the Equation (15) in \citet{oli10}. 
We applied the same procedure for SN 1999em in NGC 1637. NGC 1637 has  no TRGB distance, but its Cepheid distance is available. We adopted the Cepheid distance in \citet{leo02}, $(m-M)_0=30.34\pm0.19$. 

Derived absolute magnitudes of SN 2003gd in M74 and SN 2004dj in NGC 2403 show large differences. The $B$, $V$, and $I$-band absolute magnitudes for SN 2003gd are 0.59, 0.76, and 0.77 mag brighter than that of SN 2004dj, respectively. 
On the other hand, the differences between SN 2003gd in M74
and SN 1999em in NGC 1637 are much smaller. 
SN 2003gd is 0.22, 0.33, and 0.30 mag fainter than SN 1999em in the $B$, $V$, and $I$ bands, respectively.
Therefore  SN 2004dj in NGC 2403 might have an unusually lower luminosity than typical SNe II-P.

The Hubble constant derived from the calibration of SN 2003gd in M74 is $H_0 = 72-73\pm6$ km s$^{-1}$ Mpc$^{-1}$. It is  similar to the recent estimates based on the luminosity calibration of Type Ia supernovae \citep{rie11, fre12, lee13}. 
From the calibration of SN 1999em in NGC 1673 
we derived  $H_0 = 62-65\pm6$ km s$^{-1}$ Mpc$^{-1}$, which is in agreement with the value from
SN 2003gd within the errors.
However, since the luminosity of SN 2004dj is considerably fainter than that of SN 2003gd, 
the calibration of SN 2004dj gives a much higher value for the Hubble constant, $H_0 = 94-103\pm9$ km s$^{-1}$ Mpc$^{-1}$. 
Further studies on the luminosity calibration of more SNe II-P are needed to improve the calibration of SNe II-P.

While this paper was under review, \citet{ric14} presented $BVRI$ photometry of SN 2013ej in M74 for six months after explosion. The light curves of SN 2013ej show a clear plateau phase 5 to $\sim$90 days after discovery, and they show shapes similar to those of SN 2003gd in the same galaxy. 
The internal reddening of SN 2013ej measured from Na I D absorption lines is small ($E(B-V)=0.049\pm0.010$)\citep{ric14,val14} and so it can be an ideal target for the calibration of the SN II-P standard candle method.
Unfortunately there are no published studies for the expansion velocity of SN 2013ej at the time of day --30 from the transition phase.
When these data are available, we can use SN 2013ej
for  the calibration of the SN II-P standard candle method.

\section{Summary}

We present $VI$ photometry of the resolved stars in the arm-free region of M74 hosting three core-collapse SNe, derived from  $HST$/ACS $F555W$ and $F814W$ images. 
From the $I$-band luminosity function of the red giants, we measure the magnitude of the TRGB,
$I_{\rm TRGB}=26.13\pm0.03$, and $T_{RGB}=25.97\pm0.03$. Then we estimate the distance to M74, using the TRGB method:
$30.04 \pm 0.03 {\rm(random)} \pm 0.12 {\rm(systematic)}$
(corresponding to a linear distance of   $10.19\pm 0.14 \pm0.56$ Mpc).
Using the TRGB distance estimate to M74 in this study 
as well as the TRGB distance to NGC 2403 and the Cepheid distance to NGC 1637, %and the TRGB distance estimate to NGC 2403 hosting type II-P SN 2004dj in previous studies, 
we calibrate the standard candle method for SNe II-P.
$B$, $V$, and $I$-band absolute magnitudes for SN 2003gd are close to those of SN 1999em, while they are $\sim0.7$ mag brighter than those of SN 2004dj in NGC 2403. From these absolute magnitudes of SNe we derive a value for the Hubble constant : 
$H_0 = 72\pm 6$ (random)$\pm 7$ (systematic) km s$^{-1}$ Mpc$^{-1}$ for SN 2003gd, 
and $H_0 = 62-65\pm 6$ (random)$\pm 7$ (systematic) km s$^{-1}$ Mpc$^{-1}$ for SN 1999em. 

\bigskip

We are grateful to the referee, Barry Madore for
 helpful comments that improved the original manuscript.
This work was supported by the National Research Foundation of Korea (NRF) grant
funded by the Korea Government (MSIP) (No. 2013R1A2A2A05005120).
This paper is based on image data obtained from the Multi-mission Archive at the Space Telescope Science Institute (MAST). This research has made use of the NASA/IPAC Extragalactic Database (NED) which is operated by the Jet Propulsion Laboratory, California Institute of Technology, under contract with the National Aeronautics and Space Administration.

%\clearpage 

\clearpage

\end{document}